\documentclass[a4paper,12pt]{article}
\usepackage{graphicx}
\usepackage{amsmath}
\usepackage{amsfonts}

\begin{document}

\title{
\begin{flushright}
{\small\begin{tabular}{r} UWThPh-2001-20\\ April 2001
\end{tabular}}\\
\end{flushright}
\vspace{1cm} Magic Moments:\\ A Collaboration with John Bell}

\vspace{2cm}

\author{R.A. Bertlmann \\
Institut f\"ur Theoretische Physik, Universit\"at Wien\\
Boltzmanngasse 5, A-1090 Vienna, Austria}

\date{}

\maketitle

\begin{abstract}

I want to give an impression of the time I spent together with
John S. Bell, of the atmosphere of our collaboration and
friendship. I briefly review our work, the methods of
nonrelativistic approximations to quantum field theory for
calculating the properties of heavy quark-antiquark bound states.

\end{abstract}

\section{Prologue}

The working place of John  Bell was CERN. Near the entrance of
this huge laboratory stands the building where the Theory Division
is located. There John Stewart Bell -- often abbreviated just by
JSB -- had his office on the first floor.

It was in 1978 when I came to CERN for the first time as a young
Austrian with a Fellowship there. CERN was very impressive for its
experimental facilities, the big colliders, and as a place where
one could meet all the distinguished physicists of the field.

One of the great physicists in the Theory Division was John Bell.
He was highly respected and often consulted by his colleagues. He
could judge if a theory was right or wrong, or as he phrased it:
{\em sound or wrotten \/}. JSB was called the Oracle of CERN;
there was a certain aura around him and his office. In my first
impression the office was full of boxes where he filed the letters
and collected the works of the several fields. He himself was
sitting dignified on an armchair which sometimes dangerously bent
backwards and only he could use. In the middle of the room was
standing a double-desk, on the walls were attached two blackboards
opposite to each other. In Fig.1 you can see a little bit of his
desk.

\begin{figure}
\center{\includegraphics[width=10.9cm,
height=12.4cm]{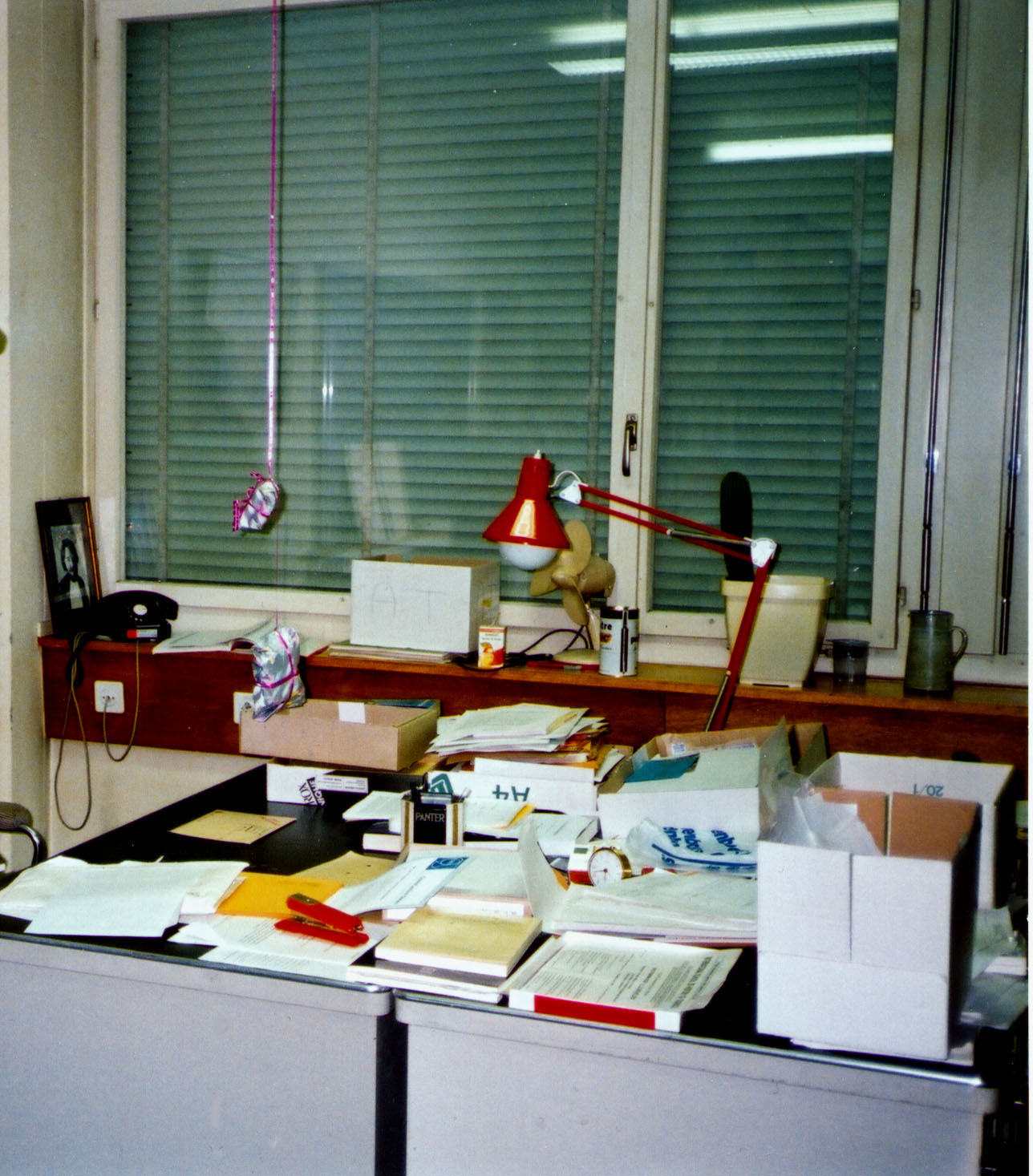}} \caption{Desk of J.S. Bell at CERN}
\end{figure}

So you can imagine how exciting it was for me to get into contact
with this man. I remember very well it was after a Seminar in the
Theory Division, we had tea in the Common Room. There he
approached me, the newcomer, and introduced himself: {\em ``I am
John Bell, where are you from? What are you working on here?"} My
interest was calculating the properties of heavy quark-antiquark
systems -- quarkonium -- specifically the wavefunctions, a hot
subject at that time since charmonium (the $J/\psi$ family) had
been found in 1974 and the even heavier system bottonium (the
$\Upsilon$ family) recently discovered in 1977.

Bell was very interested in my explanations, he even liked my
results, since they were in accordance with his Thomas-Fermi model
calculations. So our discussions in this field began and became
rather quickly a closer investigation into what is called {\em
duality}.

\section{Duality in hadronic reactions}

The conception of duality is often used in physics. What I mean is
that two seemingly different phenomena are strongly correlated to
each other; they appear as the {\em dual} aspects of one and the
same reality \cite{BertlmannActa}.\\

\vspace{0.2cm}

\noindent {\bf Hadron production in $e^+e^-$ collisions}\\

\vspace{0.02cm}

\begin{figure}
\center{\includegraphics[width=6.88cm,
height=5.55cm]{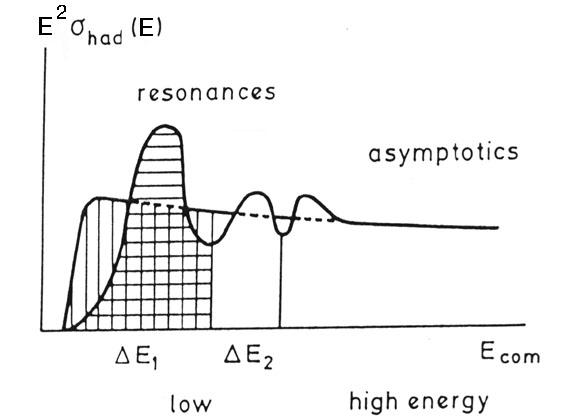}} \caption{The hadronic cross-section
in $e^+e^-$ collisions shown qualitatively}\label{fig2}
\end{figure}

\noindent Let us consider the production of hadrons (strongly
interacting particles) in electron-positron collisions. Then among
the hadrons there are also resonances produced, the $\rho, \phi,
J/\psi, \Upsilon$ families of particles
\begin{eqnarray}
e^+\;e^-&\longrightarrow\;& \textrm{hadrons}\nonumber\\ & &
\qquad\rho\;\qquad\quad \phi\;\;\;\quad J/\psi\;\quad
\Upsilon\;\dots\qquad\textrm{resonances}\nonumber\\ & &
\qquad\Downarrow\quad\qquad\; \Downarrow\quad\quad\;
\Downarrow\quad\quad  \Downarrow\nonumber\\ & &(u\bar u+ d\bar
d)\quad\; (s\bar s)\quad \;(c\bar c)\quad (b\bar b)\qquad\quad
\textrm{boundstates}
\end{eqnarray}
The cross-section for these processes I have plotted qualitatively
in Fig.\ref{fig2} and it exhibits the following feature. At low
energies some pumps appear, the resonances, but at high energies
the curve becomes quite flat or asymptotically smooth. These are
two different phenomena, the resonance production and the
asymptotic production of hadrons.

\vspace{0.2cm}

{\em Question: Are they related to each other in a dual sense?

Answer: Yes, they are.

Question: How can we understand that?}

\vspace{0.2cm}

\noindent Hadrons consist of quarks which are confined and
interact strongly via gluons (the colour force). Now, at high
energies, which corresponds to short distances, the quarks behave
as quasi-free particles. The corresponding field theory -- quantum
chromo dynamics QCD -- is asymptotically free. So there the
process
\begin{eqnarray}
e^+\; e^-\quad\longrightarrow\quad q\;\bar q
\qquad\textrm{quasi-free (anti-) quark}
\end{eqnarray}
is a good approximation. Then the cross-section approaches a
constant
\begin{eqnarray}\label{x3x}
E^2 \sigma_{q\bar q}(E)\quad\longrightarrow\quad \frac{4 \pi
\alpha^2}{3} 3 \sum_q e_q^2\;=\;\textrm{constant}
\end{eqnarray}
which is proportional to the sum of the quark charges squared
$e_q^2$ times the colour factor $3$. So we find the asymptotics
correctly.

However, at low energies, where the quarks can penetrate into
larger distances (the typical distance is of order fermi), they
are confined and generate bound states which show up as
resonances. These states are named quarkonium in analogy to
positronium.\\

\noindent {\bf Duality}\\

\noindent It was J.J. Sakurai \cite{Sakurai} who formulated
duality quantitatively. He found that if you average the
cross-section suitably over {\em all} resonances then it agrees
with the averaged asymptotics
\begin{eqnarray}\label{x4x}
\int_{\Delta E} E^2 \sigma_{res}(E) dE&=&\int_{\Delta E} E^2
\sigma_{q\bar q}(E) dE\;.
\end{eqnarray}
\medskip
Relation (\ref{x4x}) is called {\bf global duality}.

Here is the point where Bell and I began our investigations
\cite{BellBertlmannDual}.We simply asked: How far can we push
duality?\\

{\em How many resonances do we need in order to reproduce the
asymptotics?}\\

\noindent The answer is quite simple:\\ \vspace{0.2cm}

{\em One individual resonance is enough!}\\ \vspace{0.4cm}

\noindent This we called {\bf local duality}
\cite{BellBertlmannDual}
\begin{eqnarray}\label{x5x}
\Delta E\quad\longrightarrow\quad \Delta E_{res}\;.
\end{eqnarray}
The explanation why it works we found in nonrelativistic potential
theory.

\section{Nonrelativistic potential theory}

Although the bound states of quark-antiquark pairs are not ideal
nonrelativistic systems, they decay and the relative velocity of
the quarks is rather high ($v \approx 0.3 \, c$), nonrelativistic
potential theory is an amazingly powerful tool to describe their
properties.

So for our purpose of energy averaging the cross-section we can
approximate the resonances very well by a sum of delta functions

\begin{figure}
\center{\includegraphics[width=14cm, height=5.8cm]{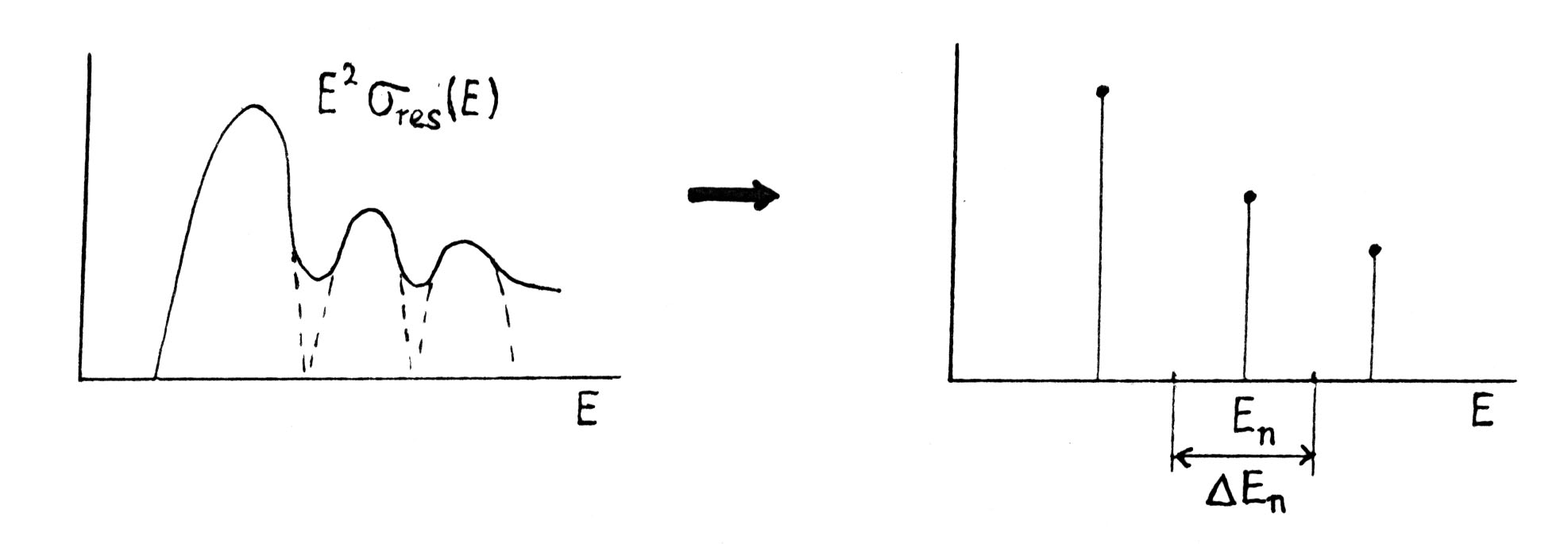}}
\caption{Approximation of resonances by delta
functions}\label{fig3}
\end{figure}
\begin{eqnarray}\label{x6x}
E^2 \sigma_{res}(E)&=&6 \pi^2 \sum_n \Gamma_n^{ee}
\delta(E-E_n)\;.
\end{eqnarray}
This replacement of the cross-section I have illustrated in
Fig.\ref{fig3}. The leptonic width of the resonance is related to
the wave function at origin by
\begin{eqnarray}\label{x7x}
\Gamma_n^{ee}&=&N \frac{|\psi_n(0)|^2}{m^2}\qquad\qquad N=\frac{4
\pi \alpha^2}{3}\cdot 3\cdot e_q^2\;.
\end{eqnarray}
On the other hand, in a nonrelativistic approximation the
cross-section for quasi-free quarks depends just on their relative
velocity $v=\sqrt{E/m}$
\begin{eqnarray}\label{x8x}
E^2 \sigma_{q\bar q}(E)&=& N \frac{3 v}{2} K(v)\;.\nonumber\\ &
&\hphantom{ N \frac{3 v}{2}}\Downarrow\nonumber\\ &
&\quad\textrm{Coulomb enhancement factor}
\end{eqnarray}
The function $K(v)$ arises from the short distance part of the
potential and is calculable. For example, in the case of the
Coulomb potential it denotes the well-known Coulomb enhancement
factor.

\begin{center}
\includegraphics[width=12.6cm, height=6.38cm]{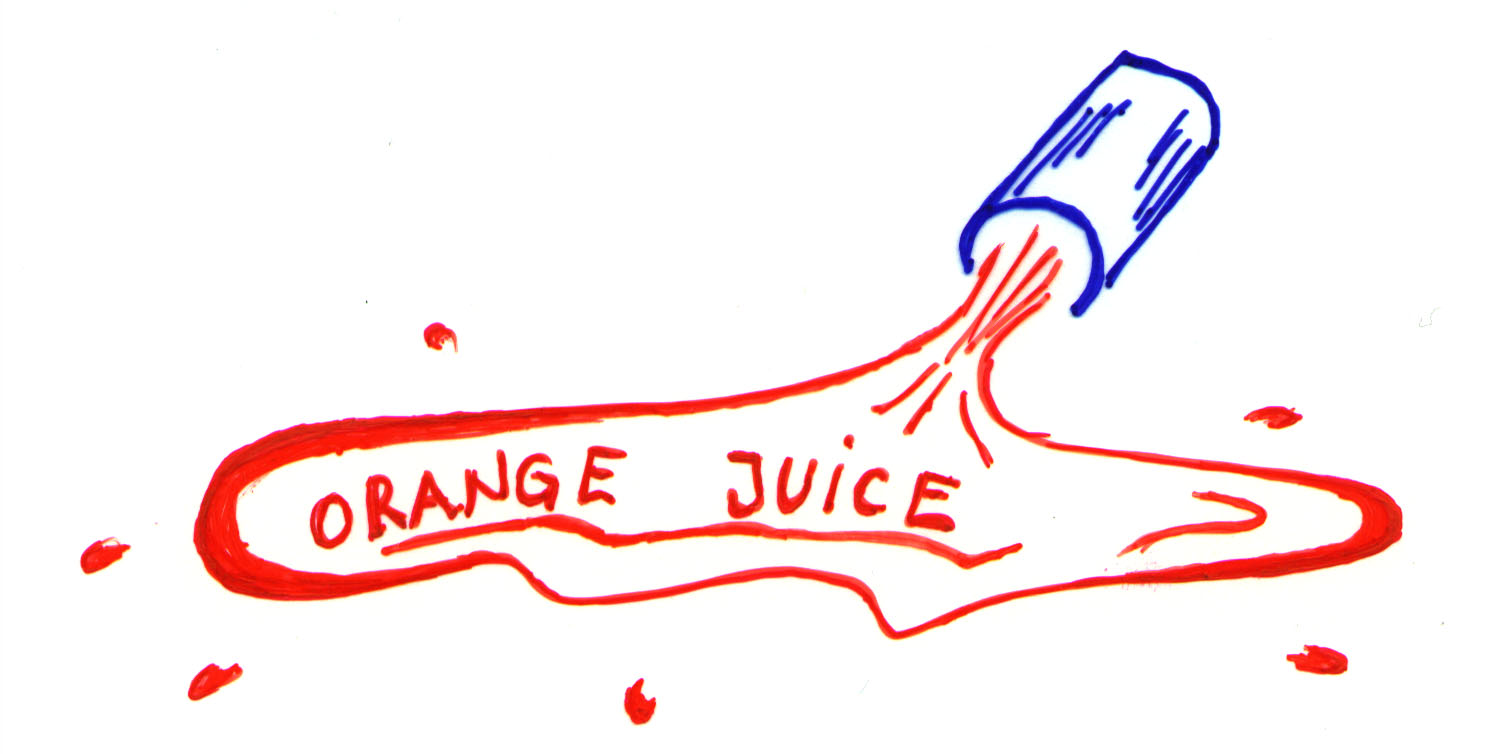}
\end{center}
At this stage I must tell a story which happened when I was
calculating some specific short distance functions $K(v)$. One day
John Bell said: {\em ``Show me your calculations at home!"} So in
the evening I went to his apartment, in a big new building in
Geneva, I took the lift to the second floor, and I was very
excited when I rang the bell. John Bell opened the door, he was
alone and offered me a drink, an orange juice. But I was so
excited that I forgot to press my fingers on the glass, the whole
orange juice dropped on the floor and polluted the beautiful
carpet. I was shocked but John smiled and said: {\em ``You sit
down and think -- I clean the floor"}. I certainly couldn't think,
I nearly fainted.\\

Coming back to expressions (\ref{x6x}) and (\ref{x8x}), if we
compare both we can check quantitatively whether duality
(\ref{x4x}) in its local form (\ref{x5x}) is true. The result for
{\bf local duality} is
\begin{eqnarray}\label{x9x}
& &4 \pi |\psi_n(0)|^2\;=\;\frac{m^2}{\pi} \int_{\Delta E_n} dE\;
v(E) K(E)\;,\nonumber\\ & &\hphantom{4 \pi
|}\Uparrow\qquad\quad\hphantom{\frac{m^2}{\pi} \int_{\Delta E_n}
dE\; v(} \Uparrow\nonumber\\ & &\qquad\quad \textrm{o.k. within
few \%}
\end{eqnarray}
where we have chosen for each resonance the energy interval
\begin{eqnarray}\label{x10x}
\Delta E\quad\longrightarrow\quad \Delta
E_n\;=\;\frac{E_n-E_{n-1}}{2}\;.
\end{eqnarray}
Calculating now the wavefunctions and the integrals in
Eq.(\ref{x9x}) explicitly it turns out that this local duality
relation holds surprisingly well. For the ground state the duality
integral in Eq.(\ref{x9x}) approximates the wavefunction within a
few percent, for the higher levels it practically coincides, and
this rather independent of the potentials we used (for details see
Ref.\cite{BellBertlmannDual}).\\ \vspace{0.05cm}

\begin{figure}
\center{\includegraphics[width=9cm, height=6cm]{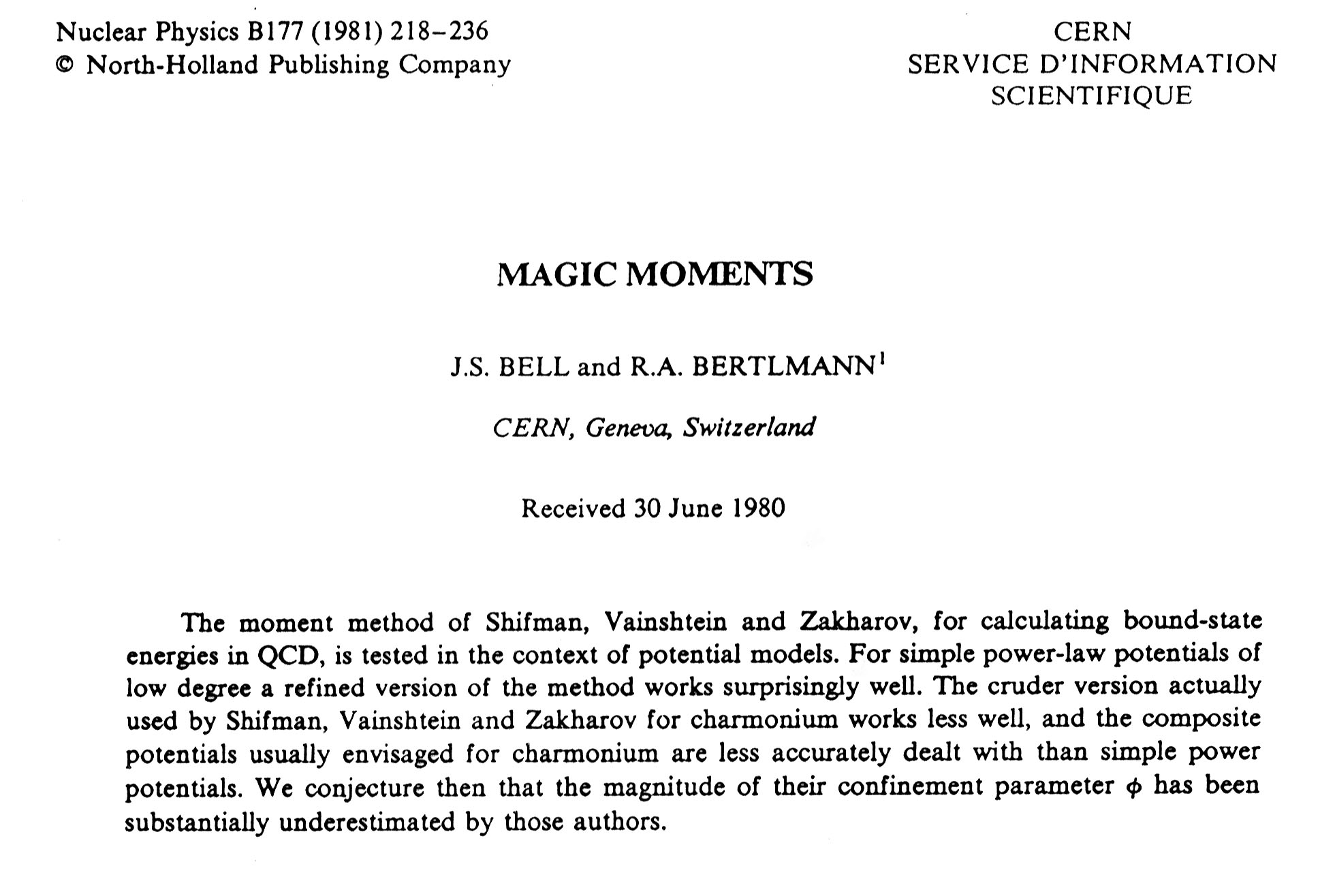}}
\caption{Paper of {\em magic moments}}\label{fig4}
\end{figure}

{\em What's the reason for it?}\\ \vspace{0.05cm}

\noindent Let us further apply the mean value theorem to the
duality integral in Eq.(\ref{x9x}) then we find an expression
which is quite familiar \cite{KrammerLealFerreira, QuiggRosner,
BellPasupathy}. It's the {\bf WKB relation}
\begin{eqnarray}\label{x11x}
4 \pi |\psi_n(0)|^2&=&
\frac{m^{\frac{3}{2}}}{\pi}\sqrt{E_n}\;K(E_n) \frac{d
E_n}{dn}\nonumber\\
\qquad\textrm{wavefunction}&\Longleftrightarrow&\qquad
\textrm{energy spectrum}
\end{eqnarray}
which relates the wavefunction at origin to the energy spectrum.
So we can trace back duality to a well-known result in quantum
mechanics.
\begin{center}
\includegraphics[width=5.74cm, height=6.54cm]{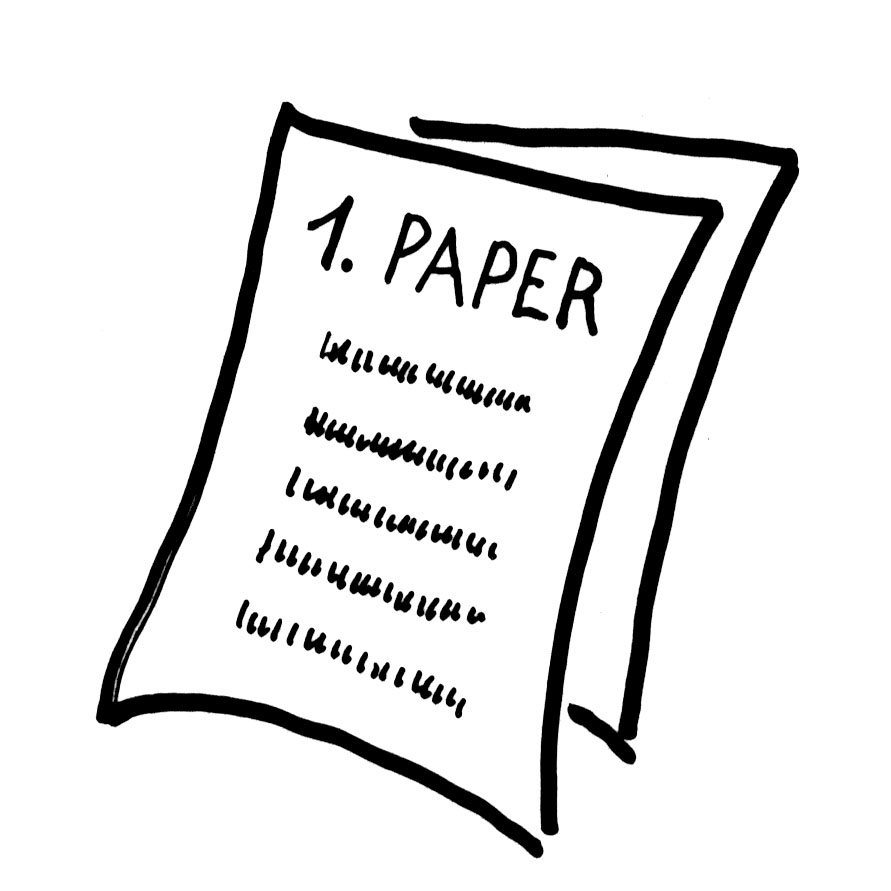}
\end{center}
When we had our results and also thought to have understood them
John proposed: {\em ``You begin to write the paper!"} I felt very
proud, worked all night, and showed him the paper the next day.
John smiled: {\em ``Oh, that looks nice, I will have a closer look
into it."} The day afterwards he returned the paper. I was shocked
to find that there was no word left in the place where I had put
it -- it was a completely different paper! But, thank God, I
slowly improved.\\ \vspace{0.01cm}

\noindent {\bf R\' esum\' e}\\ \vspace{0.005cm}

\noindent Investigating local duality, the energy smearing of a
resonance cross-section by a quasi-free quark cross-section, we
find that relation (\ref{x9x}) is very accurate independent of the
considered potential.

This feature can be understood quite naively \cite{BertlmannActa}.
In the duality relation we allow for an energy spread, which means
-- via the uncertainty relation -- that we focus on small values
of the conjugate variable time. But for short times the
corresponding wave cannot spread far enough to feel the details of
the long distance part, the confining potential. So this part can
be neglected, though certainly not the short distance part.

For this reason we can predict from a quasi-free $q \bar q$ pair
the wave function at origin of the bound state, the leptonic width
or the area of the resonance
\begin{eqnarray}\label{x12x}
\textrm{quasifree}\;q \bar q\quad\longrightarrow\quad
|\psi_n(0)|^2\;\sim\; \Gamma_{res}^{ee}\;\sim\;\textrm{area of
resonance}\;.
\end{eqnarray}
However, if we want to push the idea of duality even further in
order to become sensitive for the position of the state, the mass
of the resonance, then we penetrate into larger distances and we
must have some information about confinement.

That's the subject Bell and I got interested in next and it
resulted in a wonderful collaboration which we called {\em magic
moments \/}, see Fig.\ref{fig4}.

\section{Magic moments}

\begin{center}
\includegraphics[width=12cm, height=5cm]{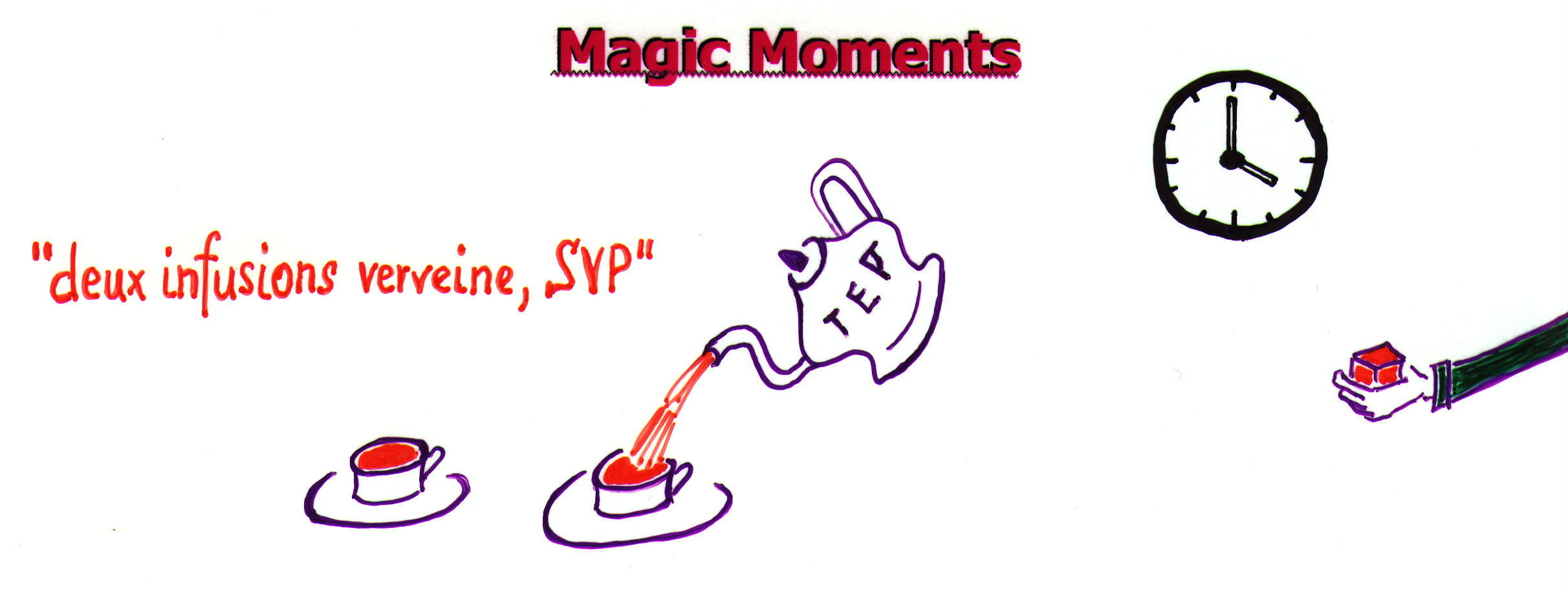}
\end{center}
One of John's habits was to have a {\em 4~o'clock tea\,}. So in
our afternoon dis\-cussions in John's office we always made a
break. It was like a ritual, at two minutes to four we left the
office and stepped down to the CERN cafeteria. There John ordered
in his typical British accent: {\em ``deux infusions verveine,
s'il vous pl\^ait"}, John's favorite tea. There, in a relaxed
atmosphere, we talked not only about physics but also about
politics, philosophy, and when Renata Bertlmann was with us we
also had heated discussions about modern art.

\begin{figure}
\center{\includegraphics[width=10cm, height=13.6cm]{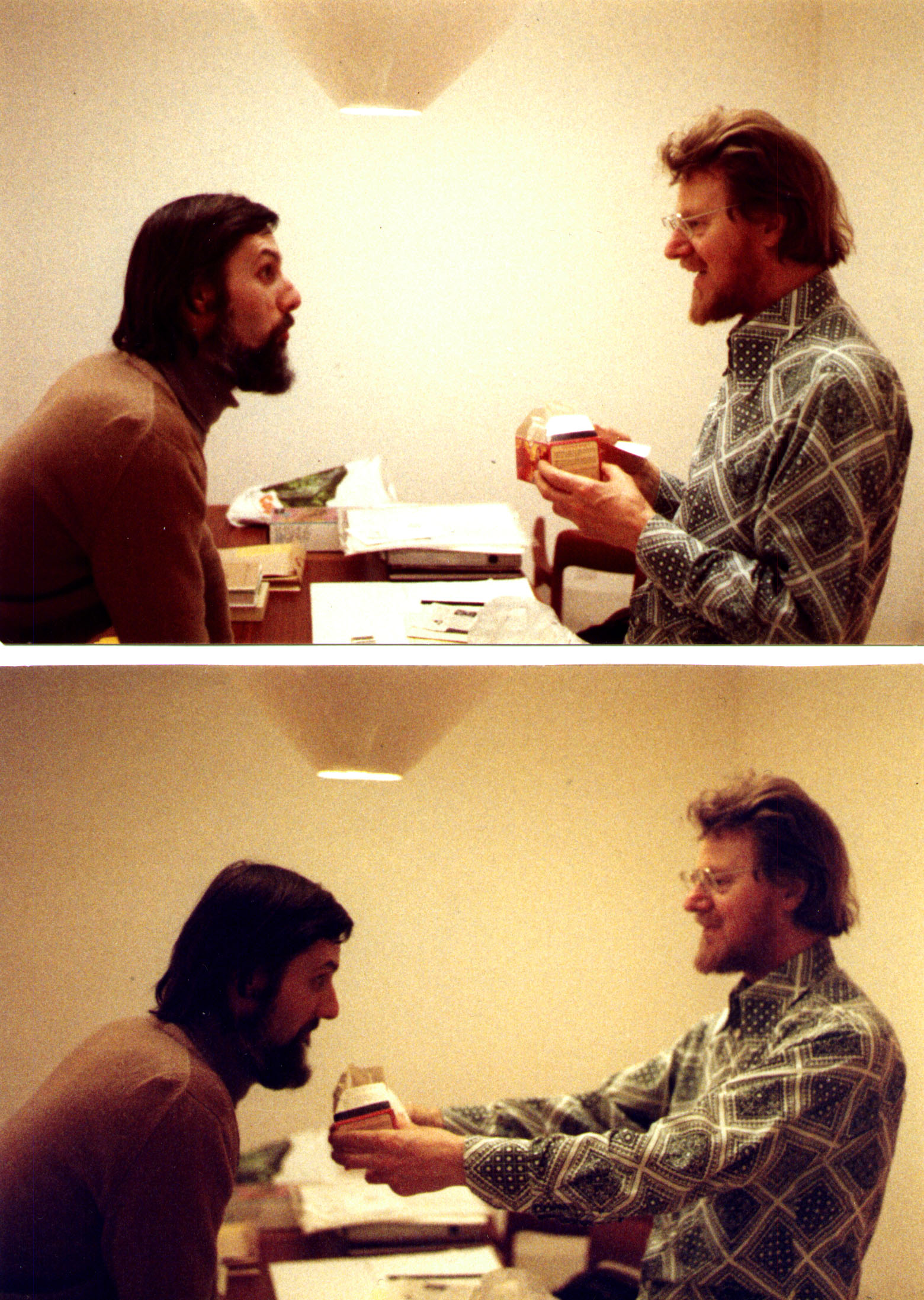}}
\caption{John Bell and Reinhold Bertlmann choosing the right sort
of tea}\label{fig5}
\end{figure}

When we worked at home a similar tea ritual took place in his
flat, and as you can see from Fig.\ref{fig5}, it took us quite a
time to choose the right sort of verveine.\\

So it was in this {\em atmosphere verveine} that John and I
discussed how to include confinement in our duality idea in order
to determine a bound state position.

\begin{figure}
\center{\includegraphics[width=8cm, height=5cm]{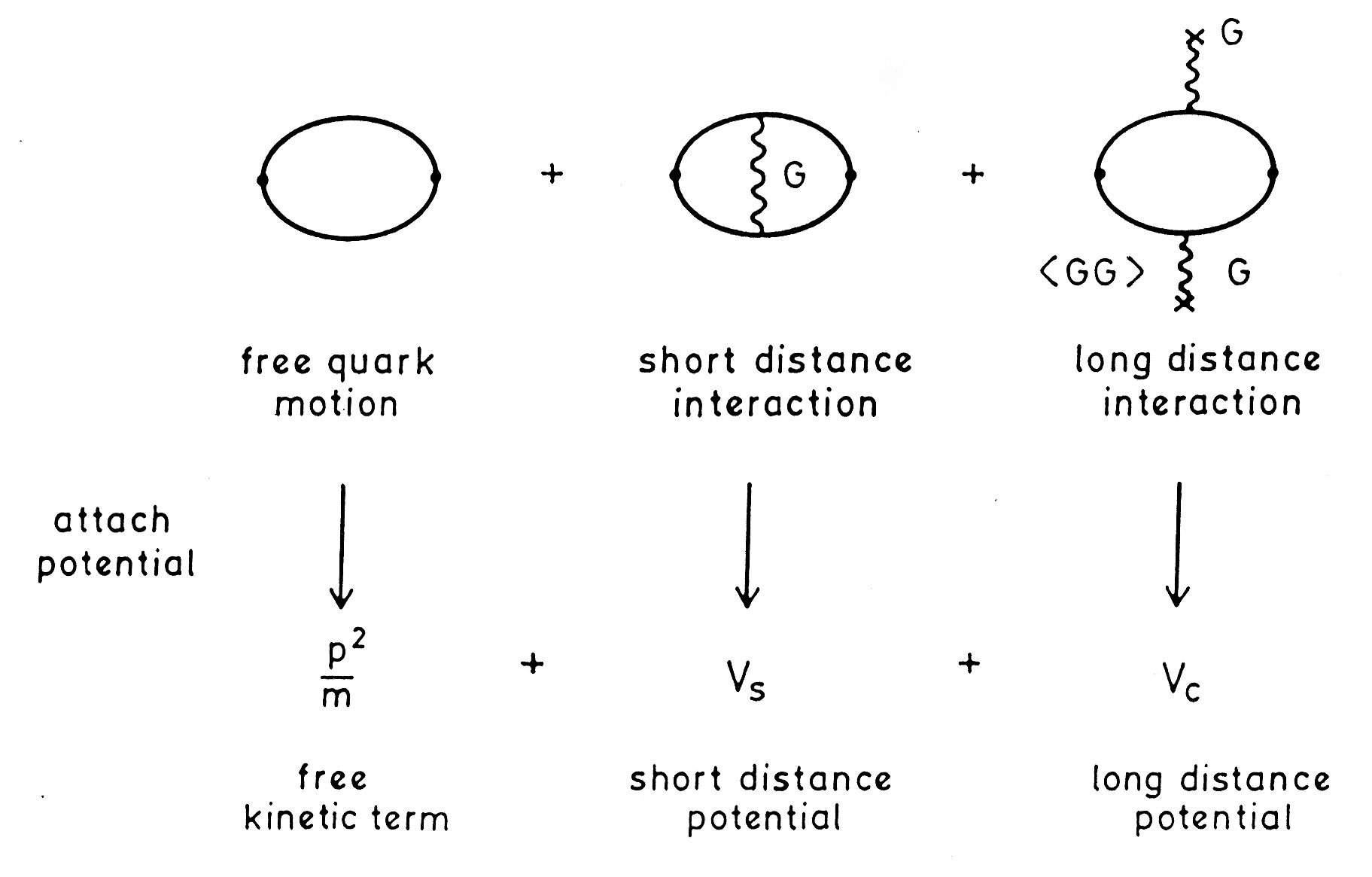}}
\caption{Feynman diagrams for the vacuum polarization
tensor}\label{fig6}
\end{figure}

Our starting point was the vacuum polarization tensor in quantum
field theory. This is the Fourier transform of the vacuum
expectation value of the time ordered product of two currents
\begin{eqnarray}\label{x13x}
i \int dx\; e^{iqx} \langle
\Omega|T\;j_\mu(x)j_\nu(0)|\Omega\rangle&=& \Pi(q^2)(q_\mu
q_\nu-q^2 g_{\mu\nu})\;.
\end{eqnarray}
It can be calculated within field theory, QCD, with help of
Feynman diagrams, loop diagrams which I have depicted in
Fig.\ref{fig6}.
The novel thing is that the usual perturbation series,
representing the short distance interaction, gets modified by
adding a small part (third diagram in Fig.\ref{fig6}), the
so-called gluon condensate $\langle GG \rangle $, the vacuum
expectation value of two gluon field strength tensors. This part
had been introduced at that time by a Russian group \cite{SVZ} in
order to account for the longer distances, the influence of
confinement. \vspace{0.2cm}

\noindent {\bf Moments}\\

\noindent Again, our investigation was
within potential theory, where we could calculate both the
perturbative {\em and} the exact result. For the energy smearing
we chose exponentials (they worked best) and we defined the
following {\bf nonrelativistic moment} \cite{BellBertlmannMagic}
\begin{eqnarray}\label{x14x}
M(\tau)&=&\int dE\; e^{-E \tau}\; Im\Pi(E)\;,
\end{eqnarray}
where $Im\Pi(E)$ is the imaginary part of the vacuum polarization
function calculable via the Feynman diagrams of Fig.\ref{fig6}.

When we do the calculation, which is actually a perturbation
theory calculation with respect to an imaginary time, the result
is the following
\begin{eqnarray}\label{x15x}
M(\tau)&=& \frac{3}{8m^2} 4 \pi
(\frac{m}{4\pi\tau})^{\frac{3}{2}}\biggl\lbrace 1+\frac{4}{3}
\alpha_S \sqrt{\pi m}\;\tau^{\frac{1}{2}}-\frac{4\pi^2}{288
m}\langle \frac{\alpha_S}{\pi} GG\rangle\;\tau^3\biggr\rbrace\;.
\end{eqnarray}
The leading term corresponds to the free motion of the quarks
(first diagram in Fig.\ref{fig6}); it is perturbed by the
$\alpha_s$ term, representing the short distance interaction
(second diagram), and by the gluon condensate $\langle GG \rangle
$ term, responsible for the longer distances (third diagram).

On the other hand, the exact moment -- which is our `experimental'
moment containing the resonances, the bound states -- is given by
the optical theorem
\begin{eqnarray}\label{x16x}
Im\Pi(E)\;\sim\;\sigma(E)
\end{eqnarray}
which relates the imaginary part of the vacuum polarization
function (the forward scattering amplitude) to the total
cross-section. Inserting this cross-section within our potential
theory, Eq.(\ref{x6x}) together with Eq.(\ref{x7x}), we get
\begin{eqnarray}\label{x17x}
Im \Pi(E)&=&\frac{3}{8m}\sum_n 4\pi |\psi_n(0)|^2 \delta(E-E_n)\;.
\end{eqnarray}
This is the expression we have to insert into Eq.(\ref{x14x}) to
find the exact nonrelativistic moment.\\ \vspace{0.1cm}

\noindent {\bf Ground state}\\ \vspace{0.05cm}

{\em How do we obtain the ground state level?}\\ \vspace{0.05cm}

\noindent To find the ground state level $E_1 \/$, the mass of the
resonance \mbox{$M = 2m + E_1 \/$}, we finally work with a
logarithmic derivative, the {\bf ratio of moments}
\begin{eqnarray}\label{x18x}
R(\tau)&=&-\frac{d}{d\tau} log M(\tau)\;=\;\frac{\int dE\; E\;
e^{-E \tau} Im \Pi(E)}{\int dE\; e^{-E \tau} Im
\Pi(E)}\quad\stackrel{\tau\longrightarrow\infty}{\longrightarrow}\quad
E_1\;.
\end{eqnarray}
For large (imaginary) times $\tau$ the ratio of moments cuts off
the contributions from the higher states and projects the ground
state energy $E_1 \/$.

In the corresponding theoretical -- perturbative -- expression we
regard the minimum value, even though not occurring at infinity,
as an approximation to $E_1$
\begin{eqnarray}\label{x19x}
& &min \mathcal{R}(\tau)\;=\;E_1\;,\nonumber\\ &
&\quad\Uparrow\qquad\qquad\quad\Uparrow\nonumber\\ &
&\;\textrm{o.k. within few \%}
\end{eqnarray}
where the theoretical ratio of moments is given by a very simple
formula
\begin{eqnarray}\label{x20x}
R(\tau)&=&\frac{3}{2\tau}-\frac{2}{3} \alpha_S \sqrt{\pi m}
\tau^{-\frac{1}{2}}+\frac{4 \pi^2}{96 m}\langle
\frac{\alpha_S}{\pi} G G\rangle \;\tau^2\;.
\end{eqnarray}
The mass of the corresponding resonance is then $M = 2m + min
R(\tau )\/$.\\ \vspace{0.05cm}

{\em What's now the result?}\\ \vspace{0.05cm}

\noindent Let me discuss a typical example, charmonium, with
values $m = 1.4 GeV$, $\alpha_s = 0.3$ and $\langle
\frac{\alpha_s}{\pi} GG \rangle = 0.02 GeV^4$. I have plotted the
ratio $R(\tau)$ in Fig.7. It shows the following typical features:

\begin{figure}
\center{\includegraphics[width=6.8cm, height=5.5cm]{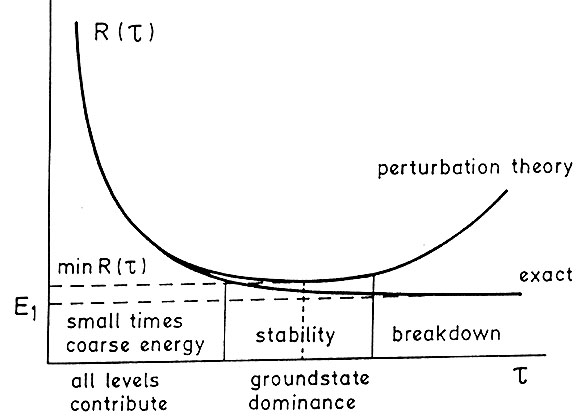}}
\caption{ratio of moments plotted qualitatively}
\end{figure}

The exact ratio approaches rather quickly its limit $E_1 \/$. The
theoretical ratio agrees perfectly for small times and stabilizes
for large times. This stability, the minimum, happens to be
already close to the ground state, quantitatively within $12\%$.
So we get a good prediction for the position of the ground state.
\vspace{0.2cm}

\noindent {\bf Balance}\\

\noindent We could show within potential theory that there exists
a balance which we can phrase in the following way
\cite{BellBertlmannMagic}:
\begin{eqnarray}
\begin{array}{l}
\textrm{course enough}\\ \textrm{pertubation theory o.k}
\end{array}\qquad&\Longleftrightarrow&
\qquad\begin{array}{l} \textrm{fine enough}\\
E_1\;\textrm{accurate}
\end{array}
\end{eqnarray}
{\em The energy average can be made coarse enough -- involving
small times -- for the modified perturbation theory to work, while
on the other hand fine enough for the individual levels to emerge
clearly.}\\

Surprising? Yes! Intuitively we had expected that for a clearly
emerging level the confinement force must be dominant and not just
a small additional perturbation. The moments, however, forced us
to re-educate our intuition, when modifying the perturbation,
levels do appear for {\em magical} reasons.

\section{Equivalent potential}

Since John and I worked within potential theory it was quite
natural for us to ask whether one can attach a potential to this
gluon condensate effect. This led us to our next collaboration
\cite{BellBertlmannPotential, BellBertlmannLett}.

\begin{figure}
\center{\includegraphics[width=12.0cm, height=8cm]{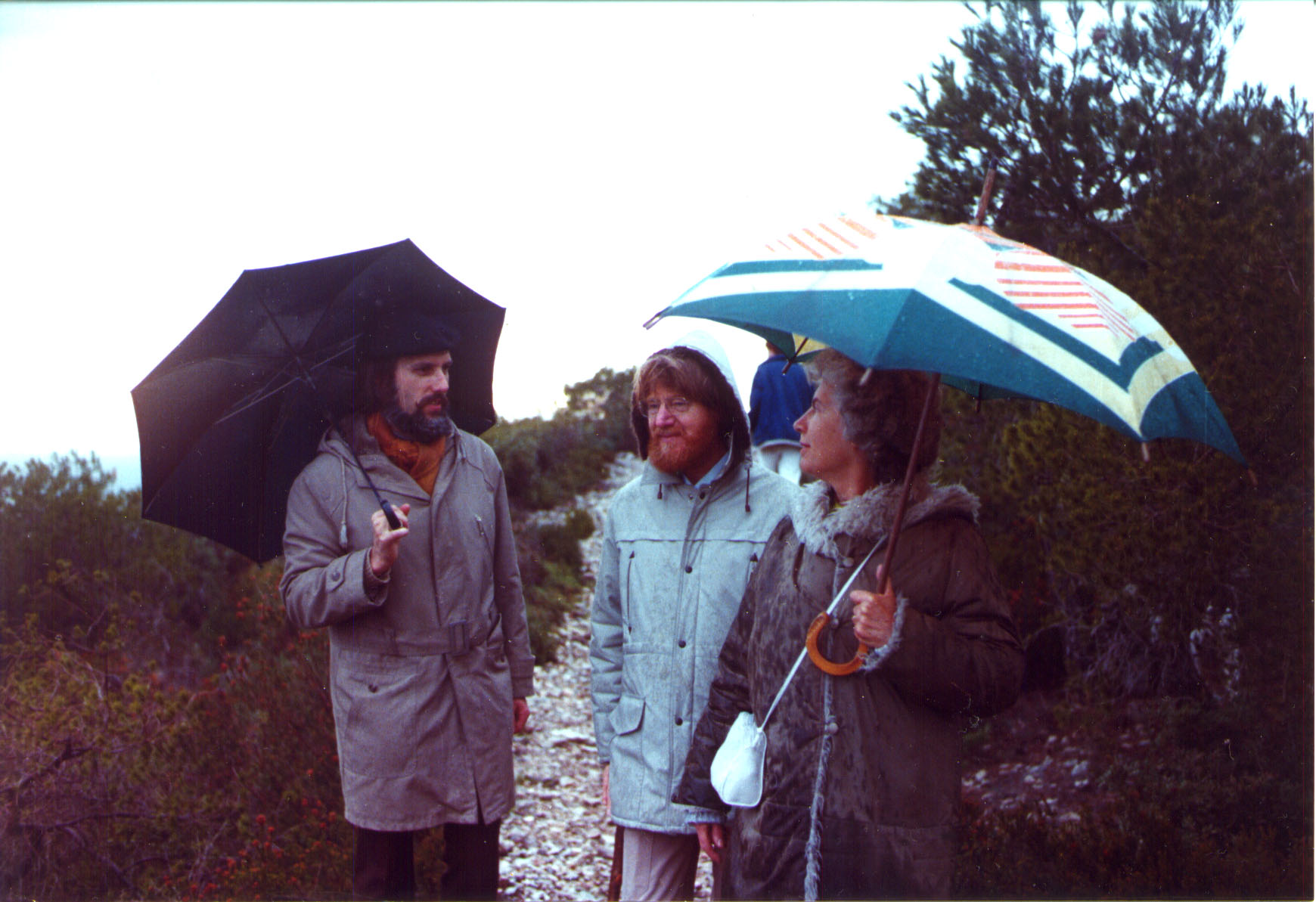}}
\caption{Reinhold Bertlmann, John Bell and Mary Bell on a walk in
the Calanque of Port Alon, South of France, 1983}\label{fig8}
\end{figure}

\begin{center}
\includegraphics[width=8.4cm, height=5.4cm]{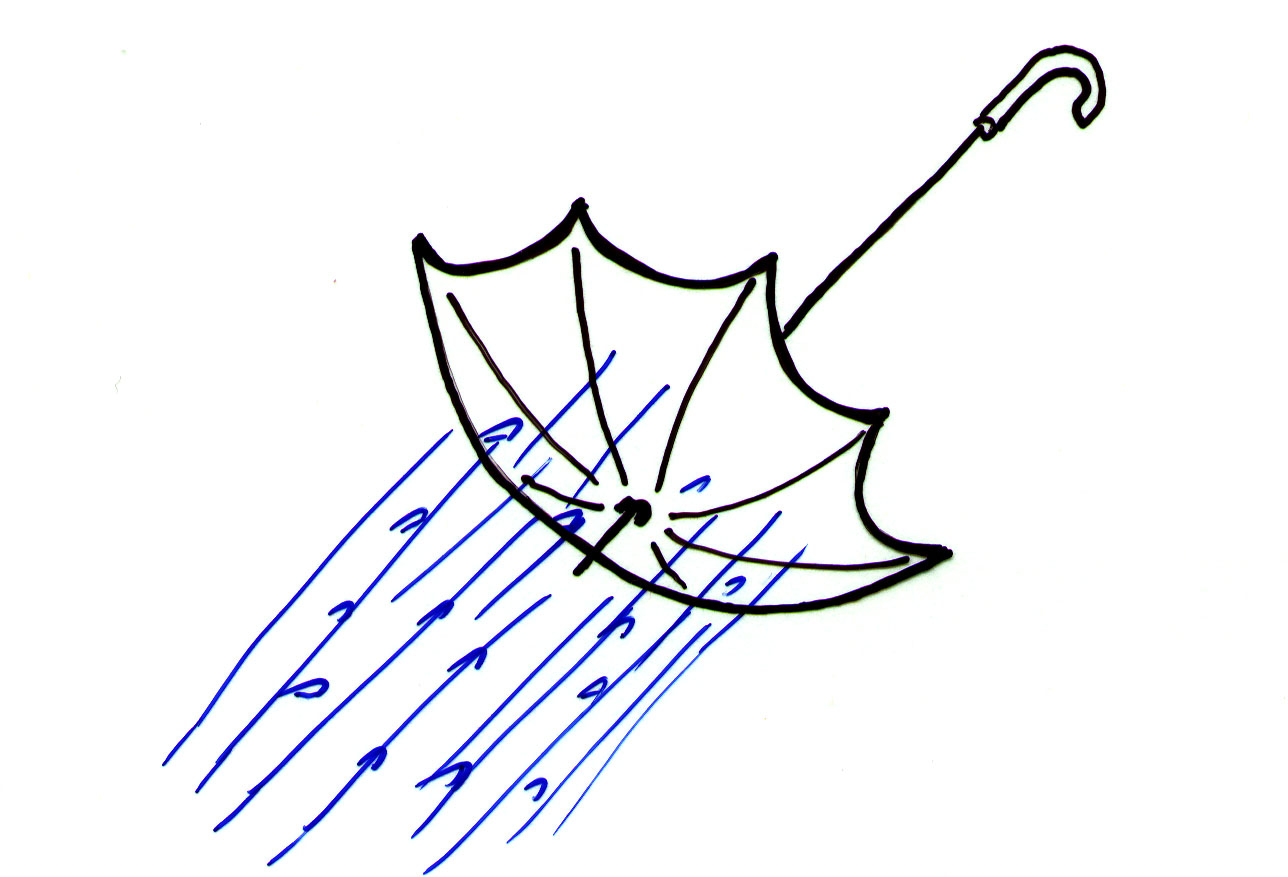}
\end{center}
One of the nice things about being a physicist is that you travel
around the world and meet your collaborators and friends
everywhere. So it happened that in the year 1983 I stayed at the
University of Marseille together with Eduardo de Rafael. He also
invited John and Mary Bell to visit the institute. There we had
interesting discussions and we also enjoyed the beautiful
surroundings. One of our walks to the Calanque of Port Alon you
can see in Fig.\ref{fig8}. We even arranged a nice British weather
for John, which was not so easy to get in the South of France.\\

When starting from quantum field theory we faced the essential
difficulty of representing the gluon condensate insertion, in the
gluon propagator, by a potential. Whereas from the short distance
part of the gluon propagator a potential can be extracted in the
usual way, the familiar Coulomb potential, the long distance part,
the gluon condensate contribution, diverges in this procedure. So
we had to look for regularized quantities. But these we had
already close at hand -- these were our {\em magic moments \/}.\\

We know already the gluon condensate effect of quantum field
theory in a nonrelativistic approximation, it is expression
(\ref{x15x}). So what we have to do is to calculate the
nonrelativistic moment (\ref{x14x}) within quantum mechanics
determined by the Hamiltonian
\begin{eqnarray}\label{x22x}
H&=&\frac{p^2}{m}+V\;,
\end{eqnarray}
where $V$ represents now the potential.

The moment defined by equation (\ref{x14x}) we can rewrite, by
virtue of expression (\ref{x17x}), as
\begin{eqnarray}\label{x23x}
M(\tau)&=&\int dE\; e^{-E \tau} Im \Pi(E)\;=\;\frac{3}{8m^2}\sum_n
4 \pi |\psi_n(0)|^2\; e^{-E_n \tau}\nonumber\\ &=&\frac{3}{8m^2}
\langle \vec{x}=0|e^{-H \tau}|\vec{x}=0\rangle
\end{eqnarray}
and we recover the (imaginary) time-dependent Green function at
$\vec x = 0\/$.

According to our previous procedure, we perturb the kinetic term
by the potential with respect to the time $\tau$
\begin{eqnarray}\label{x24x}
e^{-H\tau}&=&e^{-\frac{p^2}{m}\tau}-\int_0^\tau d\tau'\;
e^{-\frac{p^2}{m}(\tau-\tau')}\; V\; e^{-\frac{p^2}{m} \tau'}\;.
\end{eqnarray}
The perturbation integral is determined by the potential and the
familiar free Green function. For power potentials like
\begin{eqnarray}\label{x25x}
V&=&\sum_s \lambda_s r^s
\end{eqnarray}
the result for the moment is quite simple
\cite{BellBertlmannMagic}
\begin{eqnarray}\label{x26x}
M(\tau)&=&\frac{3}{8m^2} 4\pi (\frac{m}{4 \pi
\tau})^{\frac{3}{2}}\biggl\lbrace 1-\sum_s \lambda_s
\Gamma(\frac{s}{2}+1)\;m\;(\frac{\tau}{m})^{\frac{s}{2}+1}\biggr\rbrace\;.
\end{eqnarray}
This perturbation formula we have to compare with the
corresponding field-theoretic expression (\ref{x15x}). Identifying
term by term ($s =-1$ and $s = 4$) leads to the {\em equivalent
potential of Bell and Bertlmann} \cite{BellBertlmannPotential}
\begin{eqnarray}\label{x27x}
V_{BB}&=&-\frac{4\alpha_S}{3 r}+\frac{\pi^2}{144}\langle
\frac{\alpha_S}{\pi} G G\rangle\;m r^4\;.
\end{eqnarray}
It is a superposition of Coulomb and quartic potential, it is
quite steep and mass (or flavour) dependent. So it is rather
different to those favoured by the potential modellers
\cite{LuchaSchoberlGromes}.\\

Finally, John and I also considered very heavy quarkonium systems
which are very compact and dominated by the Coulomb interaction
\cite{Leutwyler, Voloshin}. There we also found a static potential
reproducing precisely the gluon condensate contribution, however,
it is also mass dependent \cite{BellBertlmannLV}.

In conclusion, no adequate bridge is found between field theory
with a gluon condensate contribution, on the one hand, and popular
potential models on the other. For an overview of this field I
refer to Ref.\cite{BertlmannPotential}.

\section{Epilogue}

John Bell was a deep and sharp thinker, a philosopher of Nature
but with the tools of a theoretical physicist. And it was his
great critical intellect that led him to his profound discoveries.
So he found together with Roman Jackiw \cite{BellJackiw} the
celebrated {\em anomalies of quantum field theory} \footnote{This
is by far Bell's most quoted paper!}, which opened the door to a
deeper understanding of Nature (see Jackiw's \cite{Jackiw}
contribution to the book), and which I could enjoy in discussing
with John \cite{Bertlmann}.\\ \vspace{0.2cm}

\noindent {Bell's famous dictum:}\\ \vspace{0.2cm}

{\em ``Speakable and unspeakable in quantum mechanics"}\\
\vspace{0.4cm}

\noindent which collects his quantum papers to a milestone book
\cite{Bell}, or its variation {\em Quantum {\rm [Un]}speakables},
the title of the conference commemorating John Bell
\cite{JSBconference}, expresses truly his character. The use of
words, their meaning, must be very precise. I could experience it
in all our works.\\ \vspace{0.2cm}

{\em Concern about terminology!}\\ \vspace{0.2cm}

\noindent This was Bell's demand. Like a moralist he attacked {\em
bad habits}, the imprecise use of words. His attacks culminated in
his brilliant article {\em Against `measurement'}
\cite{BellAgainst}, where he branded the words that should be
forbidden in any serious discussion:\\ \vspace{0.2cm}

{\em ``system, apparatus, environment, microscopic, macroscopic,
reversible, irreversible, observable, information,
measurement".}\\ \vspace{0.2cm}

\noindent For instance, {\em observables} should be replaced by
his favorite concept, the {\em beables} \cite{BellBeable}, or {\em
measurements} by {\em experiments}. But it is his great humor (for
me his typical Irish wit) which makes his attacks so delightful.
For example, \vspace{0.2cm}

{\em ``ordinary quantum mechanics is just  fine FAPP"}\\
\vspace{0.2cm}

\noindent FAPP = for all practical purposes.\\ \vspace{0.2cm}

Bell's wit you can find in all his actions, in every day life, and
also when he communicated by drawing little sketches. In one
sketch he characterized himself, and indeed, John was a gifted
cartoonist. But before I show it I have to explain it a little
bit.

\begin{figure}
\center{\includegraphics[width=6cm, height=5cm]{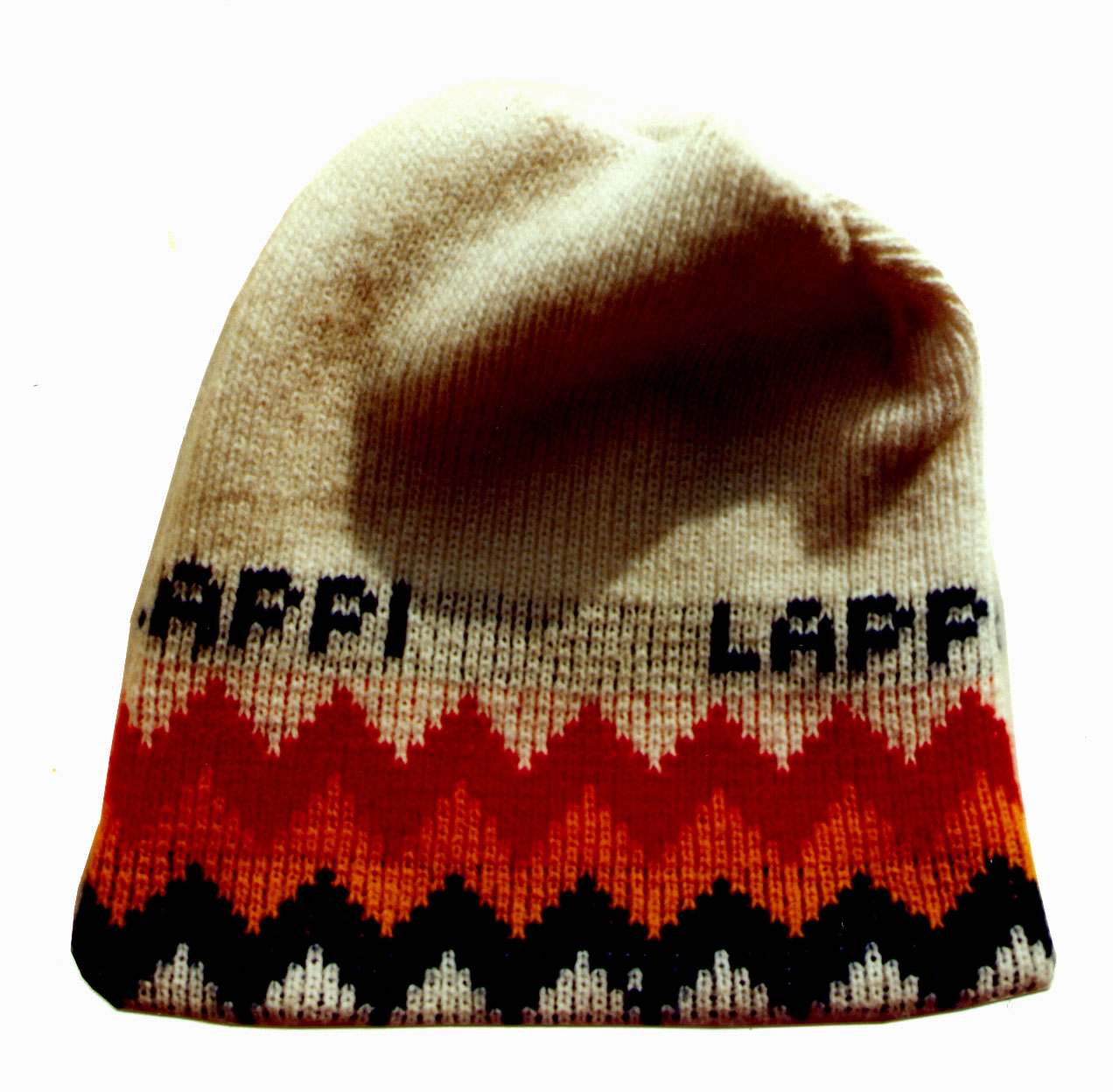}}
\caption{The {\em real} Norwegian APPI LAPPI cap of John
Bell}\label{fig9}
\end{figure}

\begin{figure}\label{fig10}
\center{\includegraphics[width=7.5cm, height=18cm]{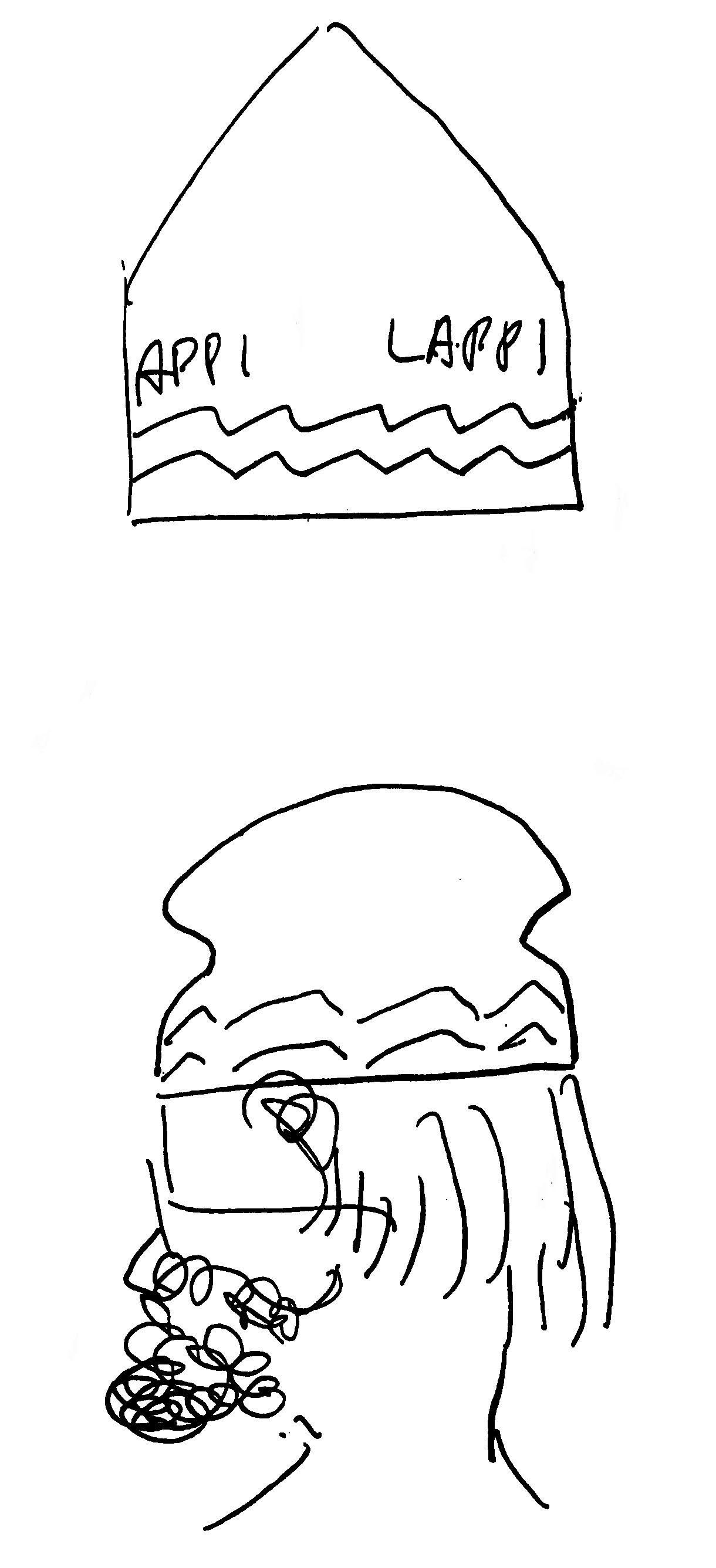}}
\caption{John Bell sketched by himself}
\end{figure}

In winter time, when it was cold, John liked to wear a Norwegian
cap. On the cap one could see the letters: APPI LAPPI embroidered.
Fig.\ref{fig9} shows the {\em real} cap.
With this APPI LAPPI cap on his head John has sketched himself --
and as you can see from Fig.\ref{fig10},
the cap, the hairs, the glasses, the beard, his image is just
perfect.\\

\begin{figure}
\center{\includegraphics[width=12cm, height=19cm]{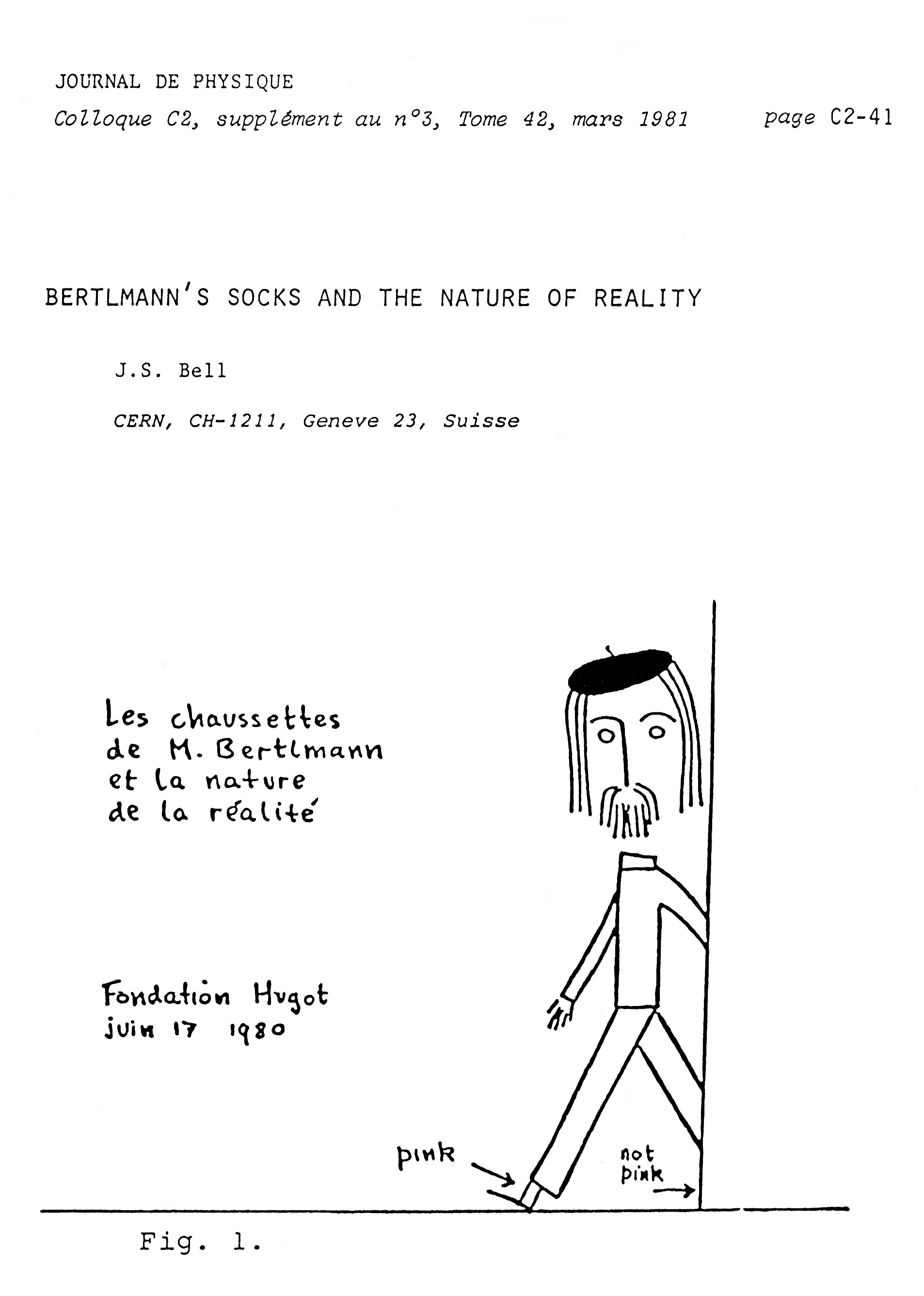}}
\caption{``Bertlmann's socks", sketch by John Bell}\label{fig11}
\end{figure}

Now I come to the story some of you might know. Also I became a
victim of his splendid wit. His article {\em Bertlmann's socks and
the nature of reality} \cite{BellBsocks}, where he described the
Einstein-Podolsky-Rosen correlations with \mbox{Bertlmann's}
socks, came out of the blue for me. In our first years of
collaboration he never mentioned his quantum works to me. And I
had not the slightest idea that he had noticed my habits of
wearing socks of different colours (a habit I had had since my
student days). Completely unexpected, this article appeared and
pushed me {\em instantaneously} into the debate of quantum
mechanics, and actually it was the cartoon, see Fig.\ref{fig11},
showing me with my odd socks in an EPR-like situation, which {\em
really} changed my life. Everybody wants to see my socks now.\\

\begin{figure}
\center{\includegraphics[width=13cm, height=17cm]{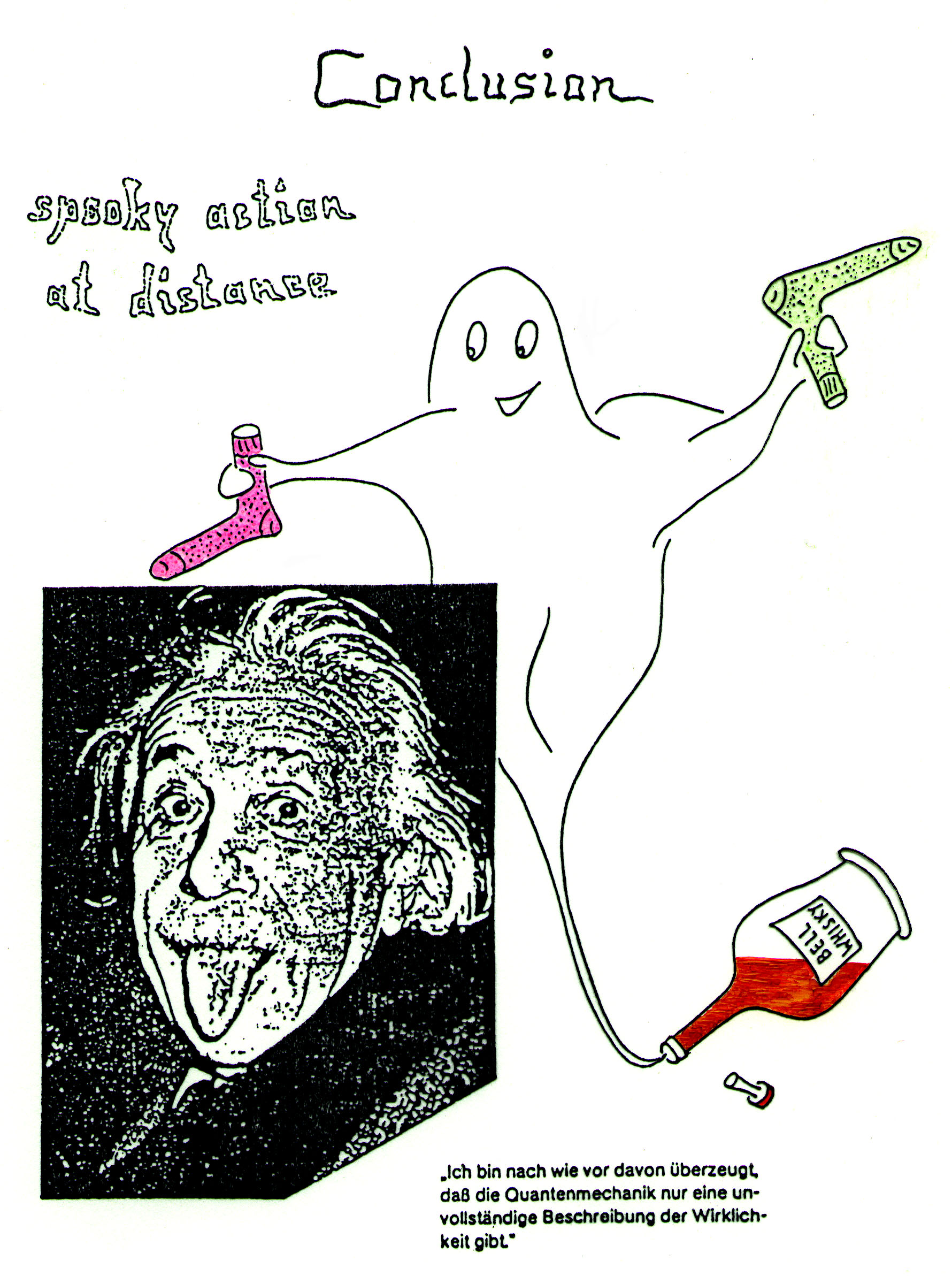}}
\caption{Conclusion of the preprint dedicated to John
Bell}\label{fig13}
\end{figure}

My answer to Bell's article was also a paper which I named {\em
Bell's theorem and the nature of reality} and which I dedicated to
him on occasion of his 60th birthday. So I wrote a preprint in
July 1988 \cite{BertlmannPreprint} and sent it to all
universities, which was the customary procedure at that time. It
appeared later on in Foundation of Physics
\cite{BertlmannBtheorem}. In the Conclusion of the paper I took a
little revenge by also drawing a cartoon, emphasizing the {\em
spooky action at a distance}, see Fig.\ref{fig13}.
It amused John very much, who was rather unaccustomed to alcohol,
that the {\em spooky action} escaped from a Bell whisky bottle
which {\em really} does exist.\\

When I recall now my collaboration with John Bell and the time we
spent together, I really can say it was a great and wonderful
time, it was {\em magic moments} indeed, and I hope I have been
able to give some impression of this in these pages.

\end{document}